# Performance-centric roadmap for building a superconducting quantum computer

R. Barends[1,*] and F.K. Wilhelm[2,†]

[1]*Institute for Functional Quantum Systems (PGI-13) and*
[2]*Institute for Quantum Computing Analytics (PGI-12),*
*Forschungszentrum Jülich, 52425 Jülich, Germany*

One of the outstanding challenges in contemporary science and technology is building a quantum computer that is useful in applications. By starting from an estimate of the algorithm success rate, we can explicitly connect gate fidelity to quantum system size targets and define a quantitative roadmap that maximizes performance while avoiding distractions. We identify four distinct phases for quantum hardware and enabling technology development. The aim is to improve performance as we scale and increase the algorithmic complexity the quantum hardware is capable of running, the algorithmic radius, towards a point that sets us up for quantum advantage with deep noisy intermediate-scale quantum computing (NISQ) as well as building a large-scale error-corrected quantum computer (QEC). Our hope is that this document contributes to shaping the discussion about the future of the field.

## I. A PERFORMANCE-DRIVEN CHALLENGE

### A. The need for performance-driven roadmapping

Quantum physics provides a potentially vast natural resource for computation. Truly tapping into that would enable quantum simulation for materials, chemistry, and high energy physics, as well as implementing algorithms related to optimization [1–5]. To bring that into reality is arguably one of the most significant technological outstanding challenges of our time. To get closer to reality, the error rate of quantum algorithms must decrease in order to permit quantum computers to grow. One should emphasize the error rate as a structuring element [6] for at least two reasons: I) Quantum computers are engineered systems, i.e. systems with external access for control and measurement, hence they cannot be kept in isolation and follow the natural trend towards classical states, known as *decoherence* [7, 8] over time. II) While equipped with binary I/O, quantum gates are controlled by analogue parameters such as angles that can be achieved by delicate control pulses and timings, hence inherit some intrinsic error sensitivity from analogue techniques. We are focusing on gate-based quantum computing; quantum annealing [9, 10] has its own set of challenges.

Here, we call for a phased roadmap where algorithm success rates are proposed as a guiding quantitative metric. The current state of the field is described, where large strides have been made in increasing qubit numbers. We would like to argue that lessons can be learned from the development of other large research infrastructure - as cutting-edge quantum computing prototypes are exactly that. From instruments such as the laser interferometer gravitational-wave observatory (LIGO) [11] and the Herschel space observatory project we can see a focus on increasing sensitivity, the need for a quantitative performance requirements map to accompany the science case, as well as the organizational approaches to dealing with pluriform consortia.

For the proposed roadmap, we identify four distinct and subsequent phases of evolution: I) identify approaches and architectures that can deliver a scalable qubit platform, II) pushing performance to a level that enables quantum advantage, III) scaling qubits and performance to deliver on quantum advantage, IV) use the developed approach and architecture to pivot to developing deep NISQ and error-corrected quantum computing.

The key element throughout this evolution is improving performance as the system size is scaled up. This implies that any developments in device physics, quantum logic gates, high quality materials, and enabling technologies contributes to the construction of a large-scale quantum computer, provided these developments are *performance-driven*.

This approach is complementary to many other roadmaps that are emphasizing qubit number [12–14]. Without the adequate increase in performance, a focus on qubit number alone can lead to systems where only a subset of the qubits can be fully entangled and thus used in beyond-classical applications. The other qubits then contribute to distracting engineering overhead. Remarkably, the EU has picked up key performance indicators [15] for their quantum computing program that are performance-driven. To quote an extreme example: The Josephson junction arrays of the 1990s [16] did already contain large numbers of elements in the quantum regime, yet, they were lacking functionality and coherence in order to count as quantum computers — our approach is complementary. Note that there has been hope that very shallow circuits can lead to some advantage in the NISQ era, which by now appears highly unlikely [17, 18].

### B. Increasing algorithmic radius

Performance can be quantified by looking at the success rate when running algorithms. See Fig. 1 for a schematic representation of a typical algorithm in its compiled form. Here, $N$ qubits undergo initialization, a set of algorithmic cycles each containing $N$ single-qubit gates and $N/2$ two-qubit entangling gates, and finally readout. Here, the two-qubit gates can be between arbitrary pairs of qubits. Arranging them in parallel layers is a measure to save time and the single qubit gates, which are either needed by the algorithm or by the compiler in order to change a native entangling gate into the desired gate [19], are assumed to be concatenated into one and then implemented as an arbitrary SU(2) rotation. The total number of cycles, or algorithm depth, is on the order of $N$ [20, 21] in order to spread entanglement across the whole qubit array and hence allow to control the full state space of the computer on the way to beyond-classical performance. There can be some overheads from limited connectivity or algorithm complexity as well as situations in which gates can be replaced by memory. Note that these type of "square" (width=depth) circuits with interleaved layers are also used in quantum volume and other volumetric performance benchmarks [22].

---

* also at: Department of Physics, RWTH Aachen University, Germany
† also at: Theoretical Physics, Saarland University, 66123 Saarbrücken, Germany

With each physical gate, error is introduced, and the success rate of the algorithm is a multiplication of fidelities of the used operations: single-qubit gates, two-qubit gates, and initialization and measurement operations. The system performance can be seen as the collection of fidelities of the elementary operations that constitute running an algorithm. The algorithm success rate can then be given by

$$F_{\text{algorithm}}(N) = F_{1q}^{A_1 N^2} F_{2q}^{\frac{1}{2} A_2 N^2} F_{\text{init}}^N F_{\text{measurement}}^N. \quad (1)$$

Here, $F$ denotes the Pauli fidelity of each operation, and $A_{1/2} \geq 1$ denote overheads for connectivity. In a generic picture, we can write $A_{1/2} = ab$ with $a$ and $b$ denoting connectivity overheads and algorithm complexity multipliers respectively. Connectivity overheads arise when gate operations between non-adjacent qubits need to be implemented, requiring SWAP operations to move the state to an adjacent qubit, and back again. In general, progress in encoding and compilation can influence these factors individually [23]. The above does not apply for constant-depth circuits, which on the other hand require large measurement and repetition overheads [24].

We have made the assumption that errors are mostly unbiased and uncorrelated, and can be approximated as single point errors. However, as correlated and biased errors can start to dominate at larger system sizes, it further underlines the need for intermediate development phases where these errors can be quantified and remedied.

The above equation illustrates that if we were to focus on increasing $N$ only, the algorithm success rate would drop exponentially. In fact, if the system performance does not improve there is no need to scale. It underlines the fundamental connection between the number of qubits to aim for and the performance of the quantum gates as implemented in the architecture. Only by improving performance $F$ and subsequently making the system size $N$ larger, can more complex quantum algorithms come within range. For a desired target algorithm success rate $F_{\text{target}}$ we can define the *algorithmic radius* of given quantum hardware and for a specific algorithm as

$$N_R = \max \{ N : F_{\text{algorithm}}(N) \geq F_{\text{target}} \}, \quad (2)$$

connecting the algorithm size, in qubit number, to the hardware system performance. This measure is related to other volumetric measures [15] with similar dependencies, and is convenient for evaluating hardware. As a guideline, we derive an upper bound to this expression for when the two-qubit gate errors dominate and are small, $\epsilon_2 \ll 1$

$$N_R \leq \sqrt{\frac{2 |\ln F_{\text{target}}|}{A_2 \epsilon_2}}, \quad \epsilon_2 = 1 - F_{2q}. \quad (3)$$

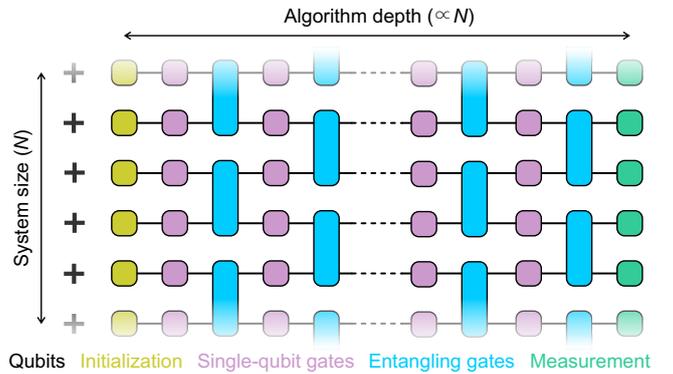

FIG. 1. Schematic of quantum algorithms such as beyond-classical demonstrations and electronic structure simulations [20, 21], with initialization, cycles of single- and two-qubit gates, and measurement at the end. Note that the two-qubit gates could be between non-adjacent qubits, which is not easy to generically draw. The algorithm success rate is a product of the fidelities of each operation, see Eq. 1.

It highlights that the precise choice of $F_{\text{target}}$ is not crucial as it only enters logarithmically. Furthermore, it shows that in order to achieve linear gain in the algorithmic radius, one should aim for quadratic reduction of gate error.

For quantitative considerations, it is still important to orient what algorithm success rate $F_{\text{target}}$ to aim for. For NP problems verification can be done in polynomial time, hence there is no lower limit on algorithm success rates in the absence of bias or correlation. Instead, the practical limits come from how many times one needs to repeat the experiment. For variational quantum eigensolving the need for a specific potential accuracy implies explicit success rates, although different error classes have distinct effects [24, 25]. As success rates depend on the exact algorithm, its implementation, as well as possible error mitigation strategies, a more practical value is needed.

Following beyond-classical demonstrations [20, 26] that have shown success rates on the order of one percent, we propose an algorithm target success rate on the order of one percent, giving an upper bound of $N_R \leq 3.0/\sqrt{A_2 \epsilon_2}$. As example, with a two-qubit error rate of 1%, assuming no extra overhead from connectivity limitations ($A_2 = 1$), $N_R = 30$. Importantly, having more qubits in the processor would neither increase algorithm success rates nor algorithm size in terms of qubit numbers given a certain target fidelity — only by improving the performance can we increase the algorithmic radius.

### C. Current state of the field — technical and structural

Since the original demonstrations in the 1980s of macroscopic quantum tunneling and energy level resonances [27–30], and the introduction of the "original three" [31–33] : charge, flux, and phase qubits around 2000, as well as the introduction of the transmon [34] (charge qubit operated in the phase regime), the field of superconducting quantum processors has developed rapidly in the last decade. This development is further underlined by achieving threshold fidelity gates on an integrated quantum chip [35] and later demonstrations of beyond-classical computing [20, 26] and below-threshold quantum error correction demonstrations [36]. The field has been reviewed at various stages [37–39].

Part of this success is based on a careful eye on device performance and fidelity. Albeit the first two-qubit gate in the superconducting platform was demonstrated as early as 2003 [40], it did not lead to an immediate aggressive scale-up but rather gradual improvements of technology towards higher fidelity, to make sure that the following first algorithm demonstrations six years later showed usable output [41].

On the other hand, millisecond coherence times have been demonstrated in isolated systems, and large-scale devices have been constructed. However, it has been an open challenge to combine both aspects: show high coherence combined with connectivity and control in large-scale systems. In such systems coherence times are now about three orders of magnitude longer than the typical fastest gate durations [20, 36, 42, 43], whereas to comfortably achieve the coherence requirements for beyond-classical algorithms and impactful error correction one needs ideally four orders of magnitude. We would like to emphasize that this number is much below the reported threshold of modern quantum error correction codes (about 0.75% for the surface code [44]). This is due to various facts: I) the threshold defines when QEC breaks even. Right below the threshold there is a transient regime and only at some distance the exponential suppression of logical errors with growing code distance can be realized. II) there is some fidelity scatter across the chip and III) the average fidelities reported on a device level average over multiple error channels whereas for QEC, some are more harmful than others, for example leakage errors out of the computational subspace [45].

Interestingly, promising alternative superconducting qubit types such as the zero-pi qubit [46–48] and fluxonium qubits [49, 50], have experimentally shown similar levels of energy coherence when considering the intrinsic quality factors $Q$. For the fluxonium the energy coherence shows a narrow peak near half a flux quantum, before dropping to a few $\mu$s. Coherence values of $T_1 = 300$ $\mu$s or even $T_1 = 1200$ $\mu$s have been shown, but only for frequencies below 1 GHz [51–53]; which can be compared with the equivalent to a $T_1$ of 60 $\mu$s at 5 GHz (as $Q = \omega T_1$, hence a lower operating frequency translates into a higher $T_1$ with equal material $Q$). For zero-pi qubits, energy coherence has been shown above 1 ms, but dephasing times vary from 8 to 26 $\mu$s [54], and how to implement high fidelity gates or rapid readout is an open topic. While the above shows that coherence has proven challenging to achieve consistently and reproduce in larger devices, it suggests that the Josephson junction itself does not inherently inhibit high levels of coherence.

At the same time, large-scale systems have been built with 72 or 1121 qubits that have not demonstrated large-scale algorithm implementations [55, 56]. Demonstrations of quantum error correction and beyond-classical benchmarking have been done with systems with frequency-tunable qubits and tunable coupling in order to achieve a high level of adressability and high-fidelity idling points of the qubits. This approach accepts opening entry points for dephasing by noise of the tuning parameters, as well as the need for an enhanced amount of control. To avoid phase decoherence associated with qubit frequency-tuning and reduce control requirements, significant effort has been placed in the construction of large scale devices using fixed coupling and microwave frequency-selective cross-resonance gates, which increase both the coherence time and the gate duration. Based on a system-level analysis, Bravyi et al. [57] recently stated that "the cross-resonance gate, which has been the core gate used in large quantum systems for the past few years, would not be the path forward".

Moreover, a lot of effort is currently focused on achieving high-fidelity two-qubit gates, even though 3- or 4-body interactions are intrinsically possible with superconducting quantum technology. Such larger body interactions can reduce gate count and hence increase algorithm success rates. On the other hand, two-qubit gates are hard enough already. We note that these interactions are more central to quantum annealing architectures, where more implementation work is delegated to hardware [58–60]

Also, different organizational approaches exist for building a quantum computer. In the United States efforts are spearheaded by multiple large self-funded IT corporations whose efforts were seeded by government investment, as well as National Quantum Information Science Research Centers. In China, a centralized government-supported effort as well as municipality-supported projects are running, that have in some cases absorbed previous industrial activities. And in Europe, nationally supported consortia as well as centralized EU-wide efforts are ongoing [6].

The above highlights that there is no single winning 'blueprint', neither for a performant large-scale control-efficient quantum processor, nor for an organizational approach, and raises the need for a critical view on how the field of superconducting qubits can optimally evolve.

## II. ROADMAP ROLE MODELS

Quantum computers of the NISQ era arguably are large research instruments: They are complex and require an infrastructure to operate, they are not disruptive in applications yet. It also becomes clear, that requirements of precision that are needed to boost up fidelities to the level that scaling becomes meaningful, are paramount and central. This is different from the challenges in classical digital computer engineering, so we look somewhere else.

### A. Analogy to gravitational wave astronomy development

As the field of quantum computing is accelerating, it is simultaneously coming up against a number of key challenges. First, current non-trivially sized systems still are not performant enough. Second, a quantitative map of what algorithms come into range as noise gets reduced and sizes scale is not well-defined yet, except for the needs for quantum error correction [61]. Third, as accurate control of large Hilbert spaces remains elusive and the exact type of architecture to build has not become apparent yet, it is becoming clear that we are entering untrodden territory.

As such, the field of superconducting quantum computing bears some resemblance to the field of gravitational-wave detectors in the 1980s. Here, resonant mass antennas were used before laser interferometers were proposed [62]. The understanding of what spectral sensitivity was needed for specific gravitational-wave sources was nascent. While the picture became more clear in the 1989 LIGO construction proposal [63] which features quantifiable milestones and a requirements map (Fig. II-2 in Ref. [63] or Fig. 5 in Ref. [64]), computing the fingerprint signal of spiralling binary black hole collisions was possible only a few years before LIGO's first detection [64]. Construction of LIGO started decades before what to look for was clearly defined — a leap of faith.

Arguably LIGO's success is thanks to the continuing improvements in detector sensitivity (see Fig. 3 in Ref. [65]), the numerical simulations in collaborations such as those started with the "Binary Black Hole Grand Challenge Alliance" [64], the organizational approach to running a long-term construction program with key milestones and followups towards the goal [66], and most importantly keeping faith in the shared vision.

While building a useful quantum computer is distinct from a gravitational-wave observatory, there are useful lessons to learn. First, the community should be challenged to provide a quantitative map on what algorithms can be achieved for specific levels of gate error and system scale. Second, continuous work on decreasing noise, in the form of gate error, needs to be a driving metric of any project. Third, a long-term vision with clear, stepped, challenging but reasonable milestones towards a shared goal is paramount to success.

### B. Performance-driven R&D

What makes building a superconducting quantum computer tough, is that it requires a broad skillset that is interdisciplinary. Constructing and operating a complex quantum system happens at the interface of architecture design, materials & fabrication, control, device physics, gates & algorithms, and software. Each of these topics comes with subtopics. Control over the quantum state, for example, requires scalable signal synthesis and delivery, as well as having a quiet qubit environment. Historically, these fields have not interacted often, one of the few examples is in the development of superconducting space instrumentation [67, 68].

In the standard view of a development cycle, new possibilities arise from novel research work done typically at universities. Here, the focus on fundamental questions about nature combines well with curiosity-driven research at low technology readiness levels (TRLs). As approaches and systems become mature and TRLs increase, the focus shifts from curiosity-driven research in academia to business-driven development in industry, encouraged by valorisation [69]. When used as a linear, one-dimensional template for building a quantum computer, this model is insufficient as it leaps over the necessary scale-up in prototype stage.

Impressive engineering efforts with a focus on scale do not necessarily yield performance, as seen in Google's Bristlecone and IBM's Condor processors [55, 56]. Moreover, a decade was needed by the Google quantum AI lab to transition from a functional 9 qubit device developed at UCSB [70] to a working 105 qubit device for subthreshold QEC demonstrations [36]. In addition, the average entangling gate error $\epsilon_2$ improved by a factor two only. Generally, pivoting corporate teams that have focused on, been built around, and incentivized for engineering, towards elucidating fundamental limitations is not straightforward [69, 71, 72].

On the other hand, in academia the focus on fundamental research in small systems, combined with the limited engineering capabilities, the need for junior scientists (graduate students and postdocs) to qualify for their next position in a limited timespan, and publish-or-perish culture arguably may make compatibility of the research work with engineering-heavy scaling efforts an afterthought.

A good example of navigating the differences amongst research groups as well as between industry while building an instrument beyond the current technical levels is a large scientific space mission, such as the Herschel Space Observatory [67]. In the end, a consortium of research labs and industry delivered a spacecraft containing three scientific instruments parked at the $L_2$ earth-sun Lagrange point. Superconductor-insulator-superconductor (SIS) receiver technology was extended beyond the superconducting energy gap frequency, stable THz local oscillators had to be developed, and novel phonon-cooled hot electron bolometer mixers, instead of the initially foreseen diffusion-cooled mixers, got into the instrument. During the development the TRL of some elements increased from concept to instrument. Acceptance of 'coopetition', where competitors and collaborators partner up and have shared responsibilities, intermediating the relationships between academia and industry, which foster different values by nature, and most importantly focus on the common science goals got it done.

A viable fast-moving quantum program draws on both advancing fundamental physics insight *and* overcoming obstacles with staunch engineering - and to bridge the gap from discovery to system. A key enabler is having a performance-driven challenge, where the goals and applications are built into the deliverables of research and development at all phases. In such an approach the strengths and weaknesses of scientists and users, academia and industry, are appreciated, and their mission aligned [67].

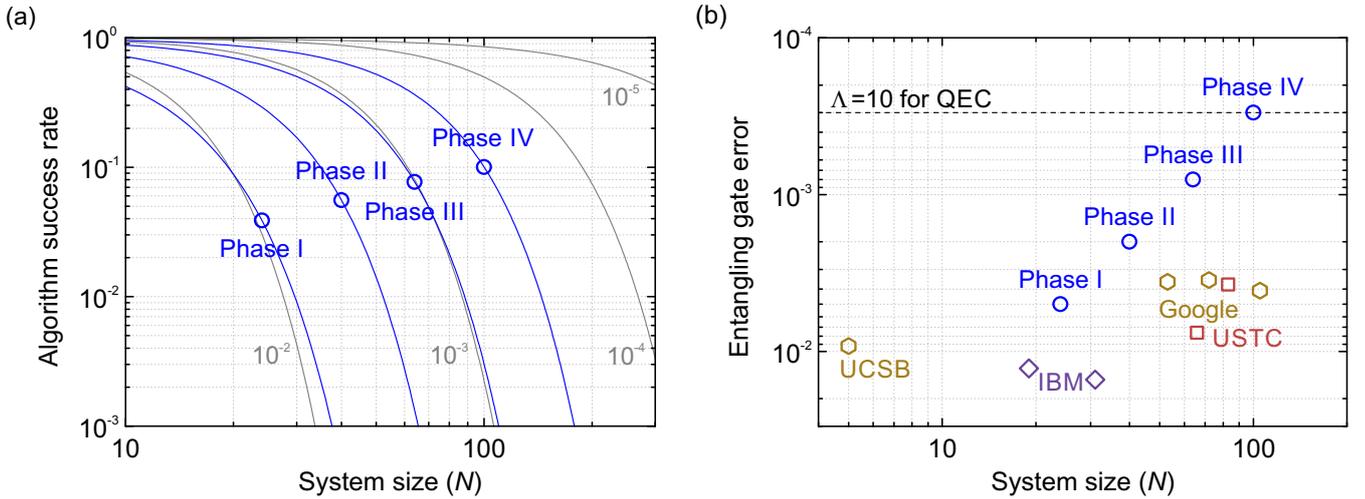

FIG. 2. (a) Estimate of the algorithm success rates for the development phase milestones in Table I as a function of the system size (number of qubits $N$) using Eq. 1 (blue). In gray, we also plot the algorithm success rate for specific entangling gate Pauli errors, here the single-qubit gate Pauli error is assumed 10% of the entangling gate error, and the combined initialization and readout Pauli error is $10^{-3}$. We set $A_1 = A_2 = 1$. The fidelity drops strongly beyond certain system sizes, underlining the connection between qubit number limits and gate fidelity metrics. (b) Comparison of entangling gate Pauli errors for the four phases and the average entangling gate Pauli error results for UCSB/Google's 5 & 9 qubit architecture [35, 70], Google Sycamore (53 qubit subset out of 54) [20] and Willow [36], IBM Hanoi (19 qubit subset out of 27) and Brooklyn (31 qubit subset out of 64) [73], and USTC Zuchongzhi-2 [26] and Zuchongzhi-3 (83 qubit subset out of 105, simultaneous application) devices [43]. In the comparison, the horizontal axis features the number of qubits in the devices used for the gate characterization, and all measured gate Pauli errors are averages of the isolated gate applications, except when noted otherwise. The requirements for quantum error correction with suppression factor $\Lambda = 10$ is dashed [61]. Note that the entangling gate fidelity target proposed for Phase II already is higher than currently demonstrated at comparable scales. When comparing gate error between distinctly different platforms, duration and parallelism of implementation can become critical parameters.

## III. DELIVERING QUANTUM HARDWARE

In developing a quantum computer, the critical challenge is maintaining quantum coherence and achieving a high level of control as we increase size as outlined in section I B. This challenge is intimately connected with the choice of architecture: one builds upon a smaller version of an architecture that works, and this isn't necessarily known a priori. Therefore we need to scale up the performance *before* scaling the system size.

In addition, to keep the pace of the development optimal, enabling technologies need to be *developed to deliver* in phases to their use, i.e., in phase(s) before their application, while keeping fully compatible with the increasing requirements of fidelity, which may contain dealing with error mechanisms that are only uncovered while the research moves along as they previously were hidden below other error sources. This need for early development of enabling technology is in particular true if one goes beyond this current roadmap and develops fully quantum error-corrected machines.

We propose development phases for superconducting quantum computers as a blueprint for heterogeneous teams as outlined in II B.

Below we discuss each phase and their goals. Performance metrics can be found in Table I. The associated algorithm success rates and comparisons with other projects can be seen in Fig. 2. The overall goal is to achieve a system with 100 qubits that can successfully run a circuit depth of over a 100 and is therefore capable of "deep NISQ", and meets the requirements for building an error-corrected quantum computer with error suppression factor $\Lambda = 10$ [61]. Such a system currently does not exist.

**Phase I: Identifying.** In the first phase the key question to answer is whether performant, scalable quantum hardware can be built. Here the deliverable is a system that can demonstrate a reasonable level of performance at a non-trivial size. The goal is to identify approaches and architectures that can deliver a scalable qubit platform - where the performance is maintained when going from a single test qubit to a system. Such a platform acts as basis for developing a large-scale quantum computer. Relevant metrics are readout, single- and two-qubit gate fidelities.

Enabling technologies related to cost-effective, scalable electronics, wiring, readout, and control need to be initiated and demonstrate a functional integration. In addition, theory work on device physics, readout, single- and two-qubit gates, electronic properties of materials, as well as benchmarking algorithms that can verify the performance of the whole system are required.

Finally, a key ingredient for performant quantum devices are coherent materials: Materials suitable for integration into large-scale superconducting systems and capable of coherent quantum transport with very low loss in the microwave regime. Developing such materials requires understanding of

TABLE I. Relevant metrics, Pauli errors, and deliverables for the technology demonstrators for each phase. Gate errors for Phase I from Ref. [74] and for Phase IV from Ref. [61]. These metrics imply algorithm success rates around a few percent following Eq. 1.

| | entangling gate error | $^2\log$(Hilbert space) | single-qubit gate error | readout & initialization error |
|---|---|---|---|---|
| Phase I Identification | 0.005 | 24 | 0.001 | 0.05 |

- Initialization & readout fidelity, single- and two-qubit gate validation using Clifford-based randomized benchmarking or cross-entropy benchmarking.
- Few-qubit quantum simulation or error correction experiments.
- Enabling technologies: Functional integration demonstrated with cost-effective, scalable electronics, wiring, readout, and control. Reproducible fabrication of coherent materials developed.

| | entangling gate error | $^2\log$(Hilbert space) | single-qubit gate error | readout & initialization error |
|---|---|---|---|---|
| Phase II Pushing performance | 0.002 | 40 | 0.0003 | 0.02 |

- System-wide benchmarking such as with cross-entropy benchmarking and other metrological tools.
- Pilot demonstrations of quantum simulation or error correction at scale.
- Enabling technologies: Integration of cost-effective electronics, dense wiring solutions, and wideband high-power amplification without performance degradation. Coherent materials able to carry performance requirements for Phase III.

| | entangling gate error | $^2\log$(Hilbert space) | single-qubit gate error | readout & initialization error |
|---|---|---|---|---|
| Phase III Showtime | 0.0008 | 64 | 0.0001 | 0.008 |

- Demonstration of viability for NISQ and QEC.
- Demonstration of large Hilbert space control.
- Deliver quantum advantage.
- Enabling technologies: Wiring and electronics cost reduced by a factor of 10. Coherent materials able to carry performance requirements for Phase IV.

| | entangling gate error | $^2\log$(Hilbert space) | single-qubit gate error | readout & initialization error |
|---|---|---|---|---|
| Phase IV Pivoting | 0.0003 | 100 | 0.00005 | 0.003 |

- Pivot to long-term trajectory for building large-scale systems for deep NISQ with error mitigation or error-corrected quantum computing.
- Demonstration of deep NISQ or $\Lambda = 10$ error suppression.
- Enabling technologies: Focus on scaling and cost reductions without performance degradation of relevant system engineering aspects.

the triangular relation between I) morphology of the superconducting thin films as base material II) microscopic insight through local or energy-resolved probing of the electron system through scanning tunneling microscopy or spectroscopic measurements, and III) macroscopic quantities like losses of quantum channels and electrical elements at high frequencies. Here, high resonator quality factors, behaving oxides, and low defect and trap counts are quantifiers. The role of good experimentation should not be overlooked. Interestingly, one of the outcomes of the search for novel materials with performance beyond the standard ones like Al in previous coherent qubit and materials programs was Al [75], but fabricated differently [76] and measured using improved experimental practices [77, 78].

Ongoing programs can act as a starting point for Phase I. The authors are currently involved in the QSOLID project, part of the German government-funded call "Quantencomputer-Demonstrationsaufbauten". Here, the deliverable is a 24 qubit device with an entangling gate error of $5 \cdot 10^{-3}$ [74]. Further metrics for this call can be found in the first column of Table I. Using Eq. 1, the algorithm success rate is targeted at $4 \cdot 10^{-2}$.

**Phase II: Pushing performance.** In the second phase this architecture is to be further developed towards improved performance at a level towards quantum advantage. The goal is to demonstrate the viability for NISQ and QEC. This is done by showing that key improvements in performance can be done and incorporated at scale, with only slightly increased qubit numbers. Relevant metrics are those using system-wide benchmarking, such as cross-entropy benchmarking, and through pilot demonstrations of quantum simulations or error correction that highlight the qubits in the architecture acting as a whole. Making use of the theory developed in the previous phase, such benchmarking can be used early in the project. This helps identifying limits in performance that may occur at scale, coming from crosstalk, parasitic coupling, and the unavoidable complexity of energy-level structures in non-trivially-sized integrated systems. We target more performant entangling gates than currently demonstrated at comparable scales (Fig. 2). Increasing the performance requires improvements in gate theory and materials. In this phase, thanks to the non-trivial size, correlated errors that can derail calcula-tions become visible and work on mitigating these will start.

Here, in preparation for the larger systems to come, enabling technologies such as cost-effective electronics, dense wiring solutions and wideband high-power amplification are to be incorporated without degradation to performance. In addition, the need for further improvements in performance will continue to encourage materials and gate development.

**Phase III: Showtime.** In the third phase, with scalability and performance demonstrated, the approach and architecture is poised for useful applications of quantum computing. Here the deliverable is controllably accessing a Hilbert space that is beyond-classical. With decreased error rates, quantum simulations that are relevant for the scientific community become possible, an important step towards commercial applications. With lower errors, the algorithm depth and therefore simulation complexity can increase. Therefore, delivering on a useful beyond-classical computation, quantum advantage, becomes a distinct reality. It is foreseeable that such a platform through providing a pathway to relevant quantum simulations can self-support part of the development that will come in the next phase thanks to commercial interest. Having developed the scalable technology in the previous phase, focus can now be placed on improving the performance and scale to where quantum advantage becomes reality.

A key technological enabler is having low-cost controls and readout. Currently, we estimate a cost of about 30 kEUR per qubit in control electronics and wiring. Arguably, these costs need to come down by at least an order of magnitude in this phase to avoid inhibiting further development. As example for room temperature solutions, with 1 GSample/sec, 14 bit digital-analog converters such as the AD9736 costing around 60 EUR each [79] that should be within reach with present-day electronics.

**Phase IV: Pivoting.** In the final phase, we take the platform having demonstrated quantum advantage and pivot to a long-term trajectory of deep NISQ as well as the construction of a large-scale quantum error-corrected system. At this point, tiling, fast error decoders, efficient control methods and electronics can become critical bottlenecks. Developments in materials, architectures, gates, and other crucial enablers should continue in a manner where advances can be straightforwardly implemented in large-scale systems.

By having focused on improving performance in the prior phases, the performance should allow for deep NISQ algorithms, using error mitigation strategies, as well as strongly sub-threshold quantum error correction. For QEC, an error suppression parameter of $\Lambda = 10$ is desirable, which becomes possible with an entangling gate Pauli error of $3 \cdot 10^{-4}$ [61]. With reproducible high performance demonstrated prior, and technology for scaling enabled, and partial self-funding thanks to deep NISQ, rapid advances are foreseeable. This final fourth phase is where the pivot to aggressive scaling can start, hitting the ground running.

In separating the development in four distinct phases, we I) maintain algorithm success rate as we scale, II) introduce the time needed for the development of architectures, materials, algorithms and importantly enabling technologies, and III) avoid the need to brute-force older methods for delivering on milestones, which is arguably unscalable and due to resource commitments may even lead to technological stagnation. Skipping these critical intermediate phases and immediately jumping to phase IV would stifle the required development and may lead to underdelivery.

## IV. CONCLUSION

We put forward a stepped plan with four phases with quantitative milestones to arrive at useful quantum computing, that is compatible with both NISQ and QEC applications. We highlight the need for a performance-driven approach that is based on algorithm success rates and provides a path towards progressively increasing the algorithmic radius. Moreover, we show that performance gains only can come from improvements in gate fidelity alongside increases in system scale.

We discuss relevant enabling technologies and at what phase their need arises. We describe the current state of the field, the pressure points between academia and industry, and challenges in arriving at scalable architectures. Moreover, we compare building a quantum computer with building an observatory, and highlight the need for a quantitative map of what algorithmic complexity can be reached for specific levels of gate error and system size.

Finally, we argue against a programmatic focus on increasing qubit numbers without first improving gate fidelity. In fact, focusing on qubit numbers while marginalizing gate fidelity improvements will lead to ever larger systems that will not stand a chance of delivering on algorithmic success — such systems have already been brute-forced. It may cause a self-amplifying loop of overpromising on qubit numbers and underdelivering on performance that has the potential to collapse the field.

Instead, a steady performance-based approach with stepped, quantifiable milestones will get us to useful quantum computing.

## ACKNOWLEDGMENTS

The authors are grateful for insightful and stimulating discussions with Teun Klapwijk, Jonas Zmuidzinas, Felix Motzoi, Stefan van Waasen, Paolo Bianco, Susanna Kirchhoff, Pavel Bushev, Yuan Gao, and Yebin Liu. We acknowledge support by the German Federal Ministry of Education and Research (BMBF), funding program "Quantum technologies - from basic research to market", project QSolid (Grant No. 13N16149).

---